# Holographic patterning of graphene-oxide films by light-driven reduction

E. Orabona[1,2,a], A. Ambrosio[1,2,3], A. Longo[4], G. Carotenuto[4], L. Nicolais[4] and P. Maddalena[1,2]

[1]*Dipartimento di Fisica, Università degli Studi di Napoli Federico II,Complesso Universitario di Monte Sant'Angelo, Via Cintia,80126, Napoli, Italy*

[2]*CNR-SPIN, U.O.S.Napoli, Complesso Universitario di Monte Sant'Angelo, Via Cintia,80126, Napoli, Italy*

[3]*School of Engineering and Applied Sciences, Harvard University, 9 Oxford Street, 02138 Cambridge, Massachusetts (USA)*

[4]*Institute for Composite and Biomedical Materials. National Research Council, 80055 Portici (NA), Italy*

a) Electronic Mail: emanuele.orabona@fisica.unina.it

We report on the patterning and reduction of graphene-oxide films by holographic lithography. Light reduction can be used to engineer low-cost graphene-based devices by performing a local conversion of insulating oxide into the conductive graphene. In this work, computer generated holograms have been exploited to realize complex graphene patterns in a single shot, differently from serial laser writing or mask-based photolithographic processes. The technique has been further improved by achieving speckle noise reduction: submicron and diffraction-limited features have been obtained. In addition we have also demonstrated that the gray-scale lithography capability can be used to obtain different reduction levels in a single exposure.

A promising method to fabricate graphene-based electronic devices is found in the reduction of Graphene Oxide (GO). GO can be easily obtained from graphite powder using the Hummers recipe[1]. The presence of oxygenated groups makes GO water-soluble and suitable for a lot of deposition strategies[2,3]. However, the same oxygen atoms break the $sp^2$ network of graphene sheets, converting the material from a conductor to an insulator. To restore the electronic properties of graphene, GO must be reduced. A large portion of oxygen-containing functional groups can be removed by chemical reducing agents[4] or by thermal treatments[5]. In 2009 Cote et al[6] reported a room temperature chemical-free reduction process where a photographic camera flash instantaneously triggers the deoxygenation reaction of GO by means of light absorption and the resulting heating. This method has allowed samples patterning by locally inducing insulator to conductor transformations by means of suitable masks into a photo-lithographic process. By this way, GO films can be used as a material in which various conductive structures can be generated on-demand to fabricate new devices. Similar results can be obtained in a maskless lithographical process by scanning a GO film with a writing laser beam, similarly to writing a CD or DVD[7]. In this process, both continuous-wave and pulsed laser writing systems have been used, obtaining small photo-reduced features[8,9], printed microcircuits[10], transistors[11] and micro-supercapacitors[12,13]. The amount of reduction and the resulting conductivity of GO have been controlled by modulating the laser power[10,14], the irradiation time[15], the number of scans[16] and the focusing condition[17]. Although effective, this lithographic process is strictly serial, resulting in long fabrication time and not uniformly overlapped areas. On the other hand, exposing a GO film to the interference pattern of multiple laser beams (Interference Lithography) allows simultaneously writing many features at once with a parallel rather than serial process. This technique has already been used to generate periodic 3D micro- and nano-scale structures on GO films[14]. However, the complexity of the shapes achievable by means of Interference Lithography is quite limited and restricted, in fact, to periodic or symmetric structures.

Holographic lithography instead, combines the main features of both mask-based and maskless approaches: it allows drawing arbitrary patterns and single features on relatively large areas, keeping micrometer resolution in a single exposure process, without any physical mask. The working principle of Holographic Lithography is to design a complex light pattern by exploiting the diffraction from a computer addressable Spatial Light Modulator (SLM). This is a Liquid Crystal on Silicon device that displays a two dimensional phase or amplitude distribution that is imposed to the incident light beam (usually working in reflection)[18,19]. SLM based holography has been shown to be suitable to produce light patterns ranging from the reproduction of real-life pictures to photonic quasicrystals with complex symmetries[20].

In this letter we apply SLM-based lithography to the patterning of GO films obtaining conductive features resulting from the light-activated reduction of GO. Furthermore, the reduction of the characteristic speckle noise affecting Holographic Lithography is proven together with the realization of different reduction grade of GO films in a single exposure.

For this experiment, GO is synthesized by applying the Hummers method to nanocrystalline graphite (i.e., graphene nano-platelets, GNP), which is prepared according to a method reported in literature[21]. In particular, GO is obtained by oxidation of 0.5g of GNP with a mixture of 25ml of sulfuric acid ($H_2SO_4$, 99.999%, Aldrich), 1g of potassium nitrate ($KNO_3$, >99.0%, Aldrich), and 3g of potassium permanganate ($KMnO_4$, >99.0%, Aldrich). Typically a mixture of GNP and $KNO_3$ in $H_2SO_4$ is stirred for few minutes below 5°C in an ice bath. $KMnO_4$ is slowly added under stirring in small portions to prevent temperature rise in excess of 20°C. Then, the temperature of the reaction mixture is raised to 35°C and the mixture stirred for 60 min. During oxidation, the color of the mixture changes from dark purplish-green to dark brown. At the end of the reaction, 100ml of water is gradually added to the solution. The suspension is further reacted by adding a mixture of $H_2O_2$ (7ml, 30%) and water (55ml).

When $H_2O_2$ solution is added, the color of the mixture changes to bright yellow, indicating that a high oxidation level is achieved by the graphite. Then, the obtained GO is separated from the reaction mixture by filtration and successively washed with water until a pH of 5-6 is obtained. The washing process is carried out by combining cycles of centrifugation with ultrasonic redispersion in water.

To obtain a thin GO film, a drop of this suspension has been cast onto a substrate and left exposed to air until the water evaporation has completed. The thickness of the obtained films is estimated to be approximately 2 µm by optical microscopy. In order to prove the reduction of GO after light exposure, Vis-NIR and Fourier transform infrared (FT-IR) spectra have been taken using respectively a Perkin Elmer Lambda 900 spectrophotometer and Spectrum 2000 FT-IR system in transmission. For FT-IR investigation the film of GO has been cast onto silicon instead of glass. The photoreduction on GO samples has been stimulated for 15 minutes by the 5mm diameter laser beam coming from a $Nd:YVO_4$ continuous-wave laser emitting at 532nm wavelength with a power of 100 mW. The visible and near infrared spectrum of GO before and after the laser light exposure is shown in Fig. 1(a). A significant absorption change in the GO film after the dehydration process is evident, in particular, the optical transmittance significantly decreases after the exposure to the laser light. The FT-IR spectrum of GO (Fig. 1(b)) before treatment (blue line) includes seven more intensive features. These features belong to vibrations of: (i) the carbon skeleton and (ii) the functional groups generated during the oxidation treatment. In particular, hydroxyl groups (-OH) and carbonyl groups (C=O) are present in the GO molecules. The -OH groups generate the C-O stretching resonance at 1072 $cm^{-1}$ and the very strong and broad stretching resonance of O-H at 3400 $cm^{-1}$. The strong carbonyl resonance absorption is visible at 1728 $cm^{-1}$, and the position of this resonance suggests the presence of esters groups. In addition to the resonances of oxygen-containing functional groups also the stretching and bending absorptions of C=C groups are well visible in the FT-IR spectrum of GO, respectively at 1621 $cm^{-1}$ and 1239 $cm^{-1}$. The absorption at 1429 $cm^{-1}$ could be attributed to the C-H bending vibration, while the stretching resonance is at 2840 cm$^{-1}$. After the laser treatment the number of resonances and their intensities decrease in the IR spectrum (red line). In particular, the C-O stretching resonance at 1072 cm$^{-1}$ completely disappears, and the C-H bending vibration at 1429 cm$^{-1}$ results strongly attenuated. The C=C stretching resonance is slightly shifted from 1621 cm$^{-1}$ to 1590 cm$^{-1}$ probably as a consequence of the conjugation extension in the graphene sheet (in fact, the absorption wavenumber is proportional to the carbon-carbon bond strength which decreases because of conjugation). The other resonances are not appreciably shifted, although their intensities are strongly decreased and a much lower signal-to-noise ratio characterizes the IR spectrum of laser-treated graphene oxide layer. According to recent literature[22,23], a thermally-induced dehydration of the GO molecules is involved in the laser treatment.

For the lithographic setup, the laser beam is expanded and sent onto a computer-controlled phase-only Spatial Light Modulator (SLM) (Pluto, Holoeye), which can be programmed for visualizing a phase map, called kinoform, producing the desired pattern. The computer-generated kinoforms, required for phase modulation, are produced by the Gerchberg–Saxton algorithm[24]. This iterative Fourier-transform-based algorithm can be used to calculate the phase distribution to be displaced onto the SLM in order to produce a predefined two-dimensional intensity distribution at the focal plane of the objective. The algorithm starts with a random phase initialization and iteratively optimizes the focal-plane intensity distribution by varying the phase values at both the image and object planes. After completing several iterations the algorithm converges into the final phase distribution. Then, the first-order diffracted beam is selected via an iris and, after recollimation, it is sent to an inverted microscope with 1.4 numerical aperture (NA) 100X objective. The optical hologram is generated at the focal plane where the GO film on a microscope glass slide is placed. Optical images can be acquired by means of a CCD camera on the microscope. Further details about the experimental setup are reported in Ref.[25].

As first test of SLM lithography on GO, the National Research Council (CNR) logo has been converted into a kinoform phase map. The resulting kinoform is shown in Fig. 2(a). Before addressing the kinoform to the SLM, a phase grating has been summed to it (convolution of Fourier Transforms), to separate the designed light pattern from the diffraction orders generated in the reflection from the SLM. This effect is due to the SLM discrete dimensions of the pixels. Fig. 2(b) shows the generated hologram in the objective focal plane that corresponds with the GO-air interface. The sample has been exposed to the hologram for 180 s with a power of 6 µW. The effects of deoxygenation process are visible in Fig. 2(c): reduced GO (r-GO) becomes dark brown in color after exposure. The reproduced logo is 17 µm × 15 µm and each line is about 1.7 µm wide. The light pattern is accurately reproduced on the sample together with the speckle distribution that affects the hologram. This is expected (see for instance Amako et al.[26]) due to the random phases algorithm initialization, that produces phase jumps between sampling points constituting the original image. These phase jumps persist even after many algorithm iterations. Then the interference of complex-amplitude fields takes place and speckle noise appears. A measure of this noise can be performed by defining a speckle contrast (C) as:

$$C = \frac{\sigma}{\mu} \tag{1}$$

where $\sigma$ and $\mu$ are the standard deviation and average of the intensity distribution calculated on the image[27]. Speckle contrast is highly dependent by the system illumination and acquiring conditions. For these reasons, the quantity defined in equation (1) cannot be used to compare images taken with different systems. However, the image contrast here defined can be still used to compare different images acquired by means of the same imaging setup. The speckle contrast value for the image reported in Fig. 2(c) is of $C_1 = 0.10 \pm 0.01$. We found out that in the lithography technique described so far, the effect of speckle on the final pattern can be strongly reduced by exploiting the statistical independence of individual kinoforms and, as a result, of the noise in this process. In fact, it can be shown that the overlapping of N statistically independent kinoforms provides an image whose speckle contrast is reduced by factor of $1/\sqrt{N}$ and it can be summarized in the following relation

$$C_N = \frac{C_1}{\sqrt{N}} \qquad (1)$$

where we have indicated with $C_N$ and $C_1$ the speckle contrast generated respectively by *N* and *1* kinoforms[28]. Alternating (sending in succession one by one) different kinoforms with a time delay of 100 ms and keeping unchanged the total exposure time, we have successfully reduced the noise in the final pattern. This is shown in Fig. 3(a) with a detail of the logo obtained alternating respectively 1, 10 and 30 independent kinoforms. The speckle contrast of these and other images has been calculated and values are reported in Fig. 3(c). The data show a behavior that cannot be simply explained by relation (2), since it does not take into account the "native" contrast of the substrate due to its roughness. Then the relation (2) should be modified as follow

$$C_N = \frac{C_1 - C_0}{\sqrt{N}} + C_0 \qquad (1)$$

where $C_0$ is the native contrast of the substrate. A fit of our experimental data with the proposed model is reported in Fig 3(c). The logo obtained with 30 kinoforms (illustrated in Fig. 3(b)) has a speckle contrast of $C_{30} = (5.5 \pm 0.3) \cdot 10^{-2}$. This value shows the improvement of the contrast obtained in designing the final pattern. Lower contrast value can be hardly reached since the best theoretical value is $C_0$; 30 kinoforms has been for us a good trade-off between kinoforms computational time and image quality. The speckle reduction allows sharpening the feature edge and increasing the homogeneity of the reduction grade in the exposed regions.

The complexity of the GO patterning by means of the technique here described can be increased by considering that an intensity modulated light pattern can be still produced by means of phase-only SLM, if a suitable image is used as reference for the Gerchberg-Saxton algorithm. In particular, if the predefined image contains pads with different gray grade, the final pattern on GO reproduces the same gray grade. Thus, by means of the technique described, it is possible to achieve different reduction grades in different zones of the GO film with a single exposure. This represents an unprecedented result that cannot be achieved by either laser scribe or interference lithography. In particular, the image of Fig. 4(a), a gray-scale with 5 different gray tones from black to white, has been converted in 30 kinoforms in order to reduce the speckle noise. The result of the lithographic process on a GO film is illustrated in Fig. 4(b) and the different visible gray levels are the proofs of different reduction grades[29]. It's important to stress again that the eight greys square (plus two blank) have been obtained simultaneously by a single exposure. The first five squares were designed with a linear decreasing (and increasing for the second row) gray-tone and, as can be verified in the Fig. 4(c), the linear trend has been preserved.

This latter experiment is particularly important since it opens new perspectives in the use of GO as an hologram recording medium exploiting the photoreduction capability to originate a variation of the refractive index[30]. Gray-scale kinoforms can be recorded and stored on GO films as already done (in a different way) with carbon nanotubes films[31]. The broad-band response of the GO and the gray scale patterning capability make GO and holographic lithography the suitable support and technique to realize graphene-based holograms. Finally, the GO build up an excellent material platform for refractive-index-based photonic applications with flexibilities, reliabilities, scalabilities and low costs.

In conclusion we demonstrated the feasibility of holographic lithography for the local transformation of GO into a graphene based material. Complex patterns have been transferred in a single shot thanks to the use of a computer generated holograms. Speckle reduction has allowed improving the quality and resolution of impressed pattern. Different reduction levels have been achieved converting a gray-scale image into a kinoform. This possibility opens the way to the exploitation of GO for photonic applications.

FIG. 1. (a) Visible, near-infrared and (b) FTIR spectra of a GO and reduced GO film.

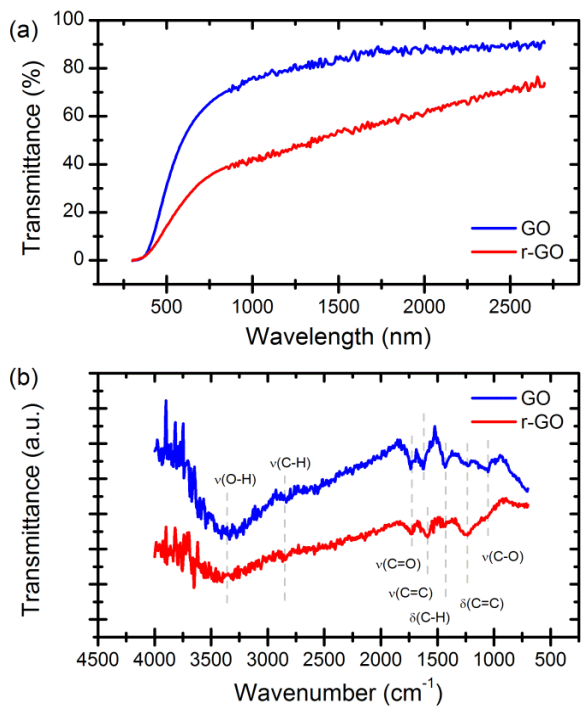

FIG. 2. (a) Calculated kinoform for the National Research Council logo. (b) Optical image of the computer generated hologram in the 100X objective focal plane. (c) Micrograph of the patterned GO film.

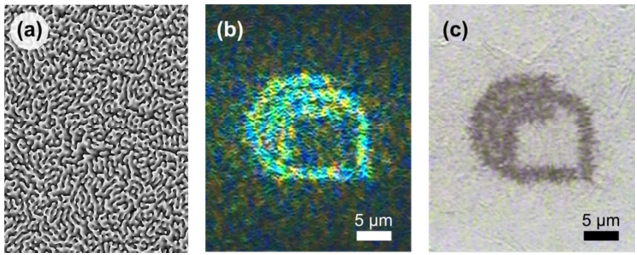

FIG. 3. (a) CNR logo and details of the impressed pattern on GO film corresponding to 1, 10 and 30 kinoforms. (b) Optical micrograph of the patterned logo obtained with 30 kinoforms; the scale bar is 5 μm long. (c) Speckle contrast as a function of kinoforms number. Points with error bars denote experimental data, while dashed line the fit performed with the proposed model.

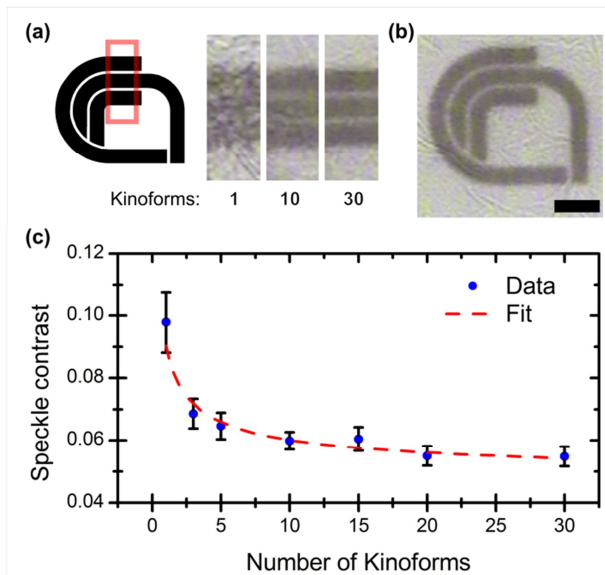

FIG. 4. (a) Designed pattern for the gray scale lithography, cyan background has been added in this image to evidence all the gray tones. (b) Optical micrograph of the impressed gray-scale pattern on to the GO film. (c) Mean gray intensity measured of the five gray tones impressed and its linear behavior.

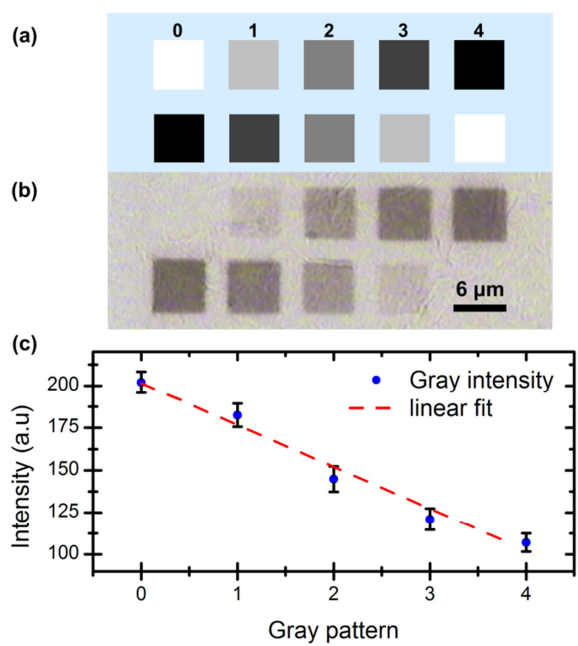